\documentstyle[preprint,aps]{revtex}
\input{epsf}
\begin{document}
\def\hatt{{\hat t}}
\def\hatx{{\hat x}}
\def\hatth{{\hat \theta}}
\def\hatta{{\hat \tau}}
\def\hatrh{{\hat \rho}}
\def\hatva{{\hat \varphi}}
\def\gsim{\mathrel{\rlap{\lower4pt\hbox{\hskip1pt$\sim$}}
    \raise1pt\hbox{$>$}}}
\def\lsim{\mathrel{\rlap{\lower4pt\hbox{\hskip1pt$\sim$}}
    \raise1pt\hbox{$<$}}}
\def\p{\partial}
\def\nn{\nonumber}
\def\tils{{\tilde s}}
\def\tila{{\tilde a}}
\def\bart{{\bar t}}
\def\barx{{\bar x}}
\def\barh{{\bar \rho}}
\def\np#1#2#3{Nucl. Phys. {\bf B#1} (#2) #3}
\def\pl#1#2#3{Phys. Lett. {\bf B#1} (#2) #3}
\def\prl#1#2#3{Phys. Rev. Lett.{\bf #1} (#2) #3}
\def\pr#1#2#3{Phys. Rev. {\bf D#1} (#2) #3}
\def\ap#1#2#3{Ann. Phys. {\bf #1} (#2) #3}
\def\prep#1#2#3{Phys. Rep. {\bf #1} (#2) #3}
\def\rmp#1#2#3{Rev. Mod. Phys. {\bf #1}}
\def\cmp#1#2#3{Comm. Math. Phys. {\bf #1} (#2) #3}
\def\mpl#1#2#3{Mod. Phys. Lett. {\bf #1} (#2) #3}
\def\cqg#1#2#3{Class. Quantum Grav. {\bf #1} (#2) #3}
\preprint{hep-th/0112106}
\title{Elliptic supertube and a Bogomol'nyi-Prasad-Sommerfield D$2$-brane--anti-D$2$-brane Pair}
\author{Jin-Ho Cho
and Phillial Oh
}
\address{{\it BK21 Physics Research Division and Institute of Basic Science\\ Sungkyunkwan University, 
Suwon 440-746, Korea\\
\sf{ jhcho@taegeug.skku.ac.kr, ploh@dirac.skku.ac.kr}
}}
\date{\today}
\maketitle
\begin{abstract}
An exact solution, in which a D$2$-brane and an anti-D$2$-brane are connected by an elliptically tubular D$2$-brane, is obtained without any junction condition. The solution is shown to preserve one quarter of the supersymmetries of the type-IIA Minkowski vacuum. We show that the configuration cannot be obtained by ``blowing-up" from some inhomogeneously D$0$-charged superstrings. The BPS bound tells us that it is rather composed of D$0$-charged D$2$-brane-anti-D$2$-brane pair and a strip of superstrings connecting them. We obtain the correction to the charges of the string end points in the constant magnetic background.
\end{abstract}
\narrowtext
The Bogomol'nyi-Prasad-Sommerfield bound states of intersecting D-branes have offered many insights on the BPS states of the world-volume quantum field theories \cite{polchinski}. Recent studies reveal that there are some systems of intersecting D-branes, including a D$0$-D$6$ bound state, which become $(1/8)$-supersymmetric when suitable $B$ fields are turned on \cite{seiberg,chen,park,witten,blumenhagen,kohta,nohta}. They could provide nontrivial models exhibiting the tachyon condensation \cite{sen}. 

On the other hand, topologically nontrivial systems with similar features have been recently reported. D$0$-charged superstrings can be ``blown-up" to a $(1/4)$-supersymmetric tubular D$2$-brane by the Poynting momentum (around the tube circumference) determined by the number densities of D$0$-branes and superstrings \cite{townsend}. This `blowing-up' is very similar to that due to the Myers effect \cite{emparan,myers}. However here, it is self-supported against collapse. This BPS configuration can also be observed in the matrix theory \cite{lee} and in the {\it T}-dual picture \cite{cho}. Even a flat D$2$-brane--anti-D$2$-brane (D$2$-$\overline{\mbox{D}2}$) pair becomes supersymmetric with D$0$-charged superstrings uniformly melted over them \cite{karch,bak}. Furthermore, supertubes with arbitrary cross section can be formed by a suitable arrangement of Born-Infeld (BI) fields over the tubes \cite{townsend2}. These results on the tubular configurations reveal that large Poynting momentum around the tube circumference yields large ``blowing-up". Can we extend this understanding to the nontubular case, expecting an abundance of the arbitrarily shaped BPS configurations of D-branes?  

The aim of this paper is to examine whether an arbitrarily shaped radius-varying (non-tubular) supersymmetric D$2$-brane can be constructed by `blowing-up' from some inhomogeneously D$0$-charged type-IIA superstrings. For this purpose, we have to consider a two-dimensional configuration, but without net D$2$-charges \cite{townsend1}. One nontrivial candidate is the D$2$/$\overline{\mbox{D}2}$ pair connected by a tubular throat. This kind of configuration was first obtained in Ref. \cite{callan} by joining two BIon solutions on a D$2$- and a $\overline{\mbox{D}2}$-brane. Although each BIon solution preserves half of the world-volume supersymmetries, (hence a quarter of the spacetime supersymmetries), the composite system is supersymmetric only when the (anti-)D-branes are infinitely far from each other making the throat extremely thin. Recalling the role of magnetic flux in the supertube case, we became puzzled about this supersymmetric configuration with no magnetic flux representing D$0$-charges over the world-volume. 

In this paper, we obtain the above nontubular configuration as an exact BPS solution (without any junction) of the Born-Infeld (BI) system describing an elliptic D$2$-brane. We show that an arbitrary amount of additional uniform magnetic fluxes over the (anti-)brane does not disturb the $(1/4)$ supersymmetries of the system. This implies that most of the physics discussed in the paper \cite{callan} are applicable to the noncommutative case. The configuration is identified as a D$0$-charged D$2$-$\overline{\mbox{D}2}$ pair and a strip of superstrings connecting the pair. We would like to emphasize that it cannot be formed by ``blowing-up" from a strip of inhomogeneously D$0$-charged superstrings.

Technically, getting our exact solution (without any junction) turns out to be possible due to the nice coordinate system we adopt for the elliptic tube. To our knowledge, this is the first exact solution describing the brane and antibrane bound system in a single coordinate system.   



We start with the elliptic tube as a simple generalization of the supertube solution found in Ref. \cite{townsend}. The extrinsic curvature on the supertube is inversely proportional to the number densities of the D$0$-branes and the superstrings melted in the tube. On the elliptic tube, the extrinsic curvature is not uniform, therefore those D$0$-branes and superstrings could be suitably distributed on the elliptic tube to make the configuration supersymmetric. In fact, this possibility is easily checked in the matrix theory context \cite{karch}. The Casimir invariant constructed from the matrix solutions of BPS equations generally describes the elliptic locus of D$0$-branes. Although the general setup of the tubular case was discussed in Ref. \cite{townsend2}, for completeness of here, we work out the elliptic tube case and obtain the explicit distribution of Poynting momentum over the elliptic tube.

The elliptic cylinder coordinates \cite{arfken} are the appropriate coordinates for our purpose, 
\begin{eqnarray}\label{coord} 
y=a\cosh{u}\,\,\cos{v},\quad z=a\sinh{u}\,\,\sin{v}.
\end{eqnarray}
Conventionally, the coordinate ranges are $u\in[0,\infty)$ and $v\in[0,2\pi)$. The ellipse is parametrized by the coordinate $v$ with the coordinate $u$ fixed.

Embedding the elliptic tube in the flat spacetime, we obtain its BI description. The metric induced on the tube reads $ds^2=-dt^2+dx^2+{\cal R}^2\left[\left(u_xdx+u_vdv \right]^2+dv^2 \right)$, where ${\cal R}^2\equiv a^2\left(\sinh^2{u}+\sin^2{v}\right)$, $u_x\equiv\partial u/\partial x$, and $u_v\equiv\partial u/\partial v$, and $\{t,x,v\}$ constitute the world-volume coordinates. We will consider only the elliptic configuration, so we set $u_v=0$. Assuming the following time-independent BI field strengths, $F=E\,dt\wedge dx+B\,dx\wedge dv$, we construct the BI Lagrangian,
\begin{eqnarray}\label{lag}
{\cal L}=-\sqrt{{\cal R}^2\left(1-E^2\right)+{\cal R}^4u_x^2+B^2}.
\end{eqnarray}

For a given canonical momentum $\Pi=\partial{\cal L}/\partial E=-E{\cal R}^2/{\cal L}$ and the magnetic field $B=B(v,x)$, we are looking for the coordinate $u$ which minimizes the Hamiltonian,
\begin{eqnarray}\label{hamiltonian}
{\cal H}=\sqrt{\frac{\left({\cal R}^2+{\cal R}^4u_x^2+B^2\right)\left({\cal R}^2+\Pi^2\right)}{{\cal R}^2}}.
\end{eqnarray}
For the tubular case of $u_x=0$, it is minimized when $a^2\left(\sinh^2{u}+\sin^2{v}\right)=\vert B\Pi\vert$. If $B\Pi$ is independent of the coordinate $v$, this result is inconsistent with our starting assumption of $u_v=0$. The only way to obtain the consistency is if $\vert B\Pi\vert=C^2+a^2\sin^2{v}$ with some constant $C$. The coordinate $u_0$ minimizing the Hamiltonian is determined by the relation, $a^2\sinh^2{u_0}=C^2$. At this point, the Hamiltonian is saturated to the BPS bound; ${\cal H}=\vert B\vert+\vert\Pi\vert$. This implies that the elliptic tube consists of the D$0$-branes and the superstrings whose number densities are given by $\vert B\vert$ and $\vert \Pi\vert$, respectively. The electric field at $u=u_0$ becomes, $E=\mbox{sgn}(\Pi)=\pm1$ as in the case of the circular supertube \cite{townsend}.

Let us check the equations of motion: 
\begin{eqnarray}\label{eom}
\frac{\partial\Pi}{\partial x}=0;\quad
\frac{\partial}{\partial x}\left(\frac{B}{{\cal L}}\right)=\frac{\partial}{\partial v}\left(\frac{B}{{\cal L}}\right)=0.
\end{eqnarray}
The former equation is the Gauss law constraint, which tells us in the tubular case ($u_x=0$) that the density of  superstrings is uniform along the axial direction of the tube. The above solutions minimizing the Hamiltonian satisfy the remaining equations of motion, as long as the field $B$ does not change its sign over the tube.


Since the above elliptic solution is stable and is a BPS solution, it is reasonable to expect the solution to preserve some supersymmetries. In the following, we show that this is true. The above elliptic tube is $(1/4)$ supersymmetric. Supersymmetry is determined by solving the Killing spinors equation, $\Gamma\epsilon=\epsilon$, where $\Gamma$ is the matrix defining $\kappa$ transformation on the world-volume of D-branes \cite{bergshoeff} and is given in our case as
\begin{eqnarray} 
-{\cal L}\Gamma=-{\cal R}\Gamma_{023} +{\cal R}^2u_x\Gamma_{012}+E{\cal R}\Gamma_{2}+B\Gamma_0\Gamma_{\natural}.
\end{eqnarray}
Here $\{\Gamma_i: i=0,1,\cdots,9\}$ are the spacetime Dirac matrices, $\Gamma_{\natural}=\prod\limits_{i=0}^{9}\Gamma_i$, and we set $\{x^0,x^1,x^2,x^3\}=\{t,y,z,x\}$. 

In order to solve the Killing spinor equation, it is convenient to use the covariantly constant spinor $\epsilon=\exp[{-\frac{i}{2}\tan^{-1}\left({\coth{u}\tan{v}}\right)\Gamma_{12}}]\epsilon_0$ adequate to the elliptic coordinate frame. Inserting this expression into the Killing spinor equation yields the following two conditions on the constant $32$-component spinor $\epsilon_0$;
\begin{eqnarray}\label{gamma}
&&\left(\Gamma_{03}\Gamma_{\natural}+E\right)\epsilon_0=0,\nonumber\\
&&\left(B\Gamma_0\Gamma_{\natural}+{\cal R}^2u_x\Gamma_{012}+{\cal L}\right)\epsilon_0=0.
\end{eqnarray}
The first equation tells us that $E^2=1$. Although this result was already obtained for the tubular case, we stress here that it applies to the more ``general'' case of $u_x\ne 0$. With this condition, the second equation reduces to
\begin{eqnarray}
\left({\cal R}^2u_x\Gamma_{012}+B\Gamma_0\Gamma_{\natural}\right)\epsilon_0
=\sqrt{{\cal R}^4u_x^2+B^2}\,\,\epsilon_0.
\end{eqnarray}

When $u_x=0$, the above equation is further simplified as $\Gamma_0\Gamma_{\natural}\epsilon_0=\mbox{sgn}(B)\epsilon_0$. Since $\Gamma_0\Gamma_{\natural}$ is a product structure (it is squared to the identity and is traceless) commuting with the other product structure $\Gamma_{03}\Gamma_{\natural}$ concerning the field $E$, $(1/4)$ supersymmetries are preserved upon the imposition of the conditions (\ref{gamma}).

When $u_x\ne 0$, we let $\cos{\alpha}\equiv B/\sqrt{{\cal R}^4u_x^2+B^2}$.
Then the second Killing spinor condition becomes $\left(\sin{\alpha}\,\Gamma_{012}+\cos{\alpha}\,\Gamma_0\Gamma_{\natural}\right)\epsilon_0=\epsilon_0$. Especially when $B=B_0{\cal R}^2u_x$ (including the case of $B_0=0$), the operator $\sin{\alpha}\,\Gamma_{012}+\cos{\alpha}\,\Gamma_0\Gamma_{\natural}$ becomes a coordinate-independent product structure commuting with $\Gamma_{03}\Gamma_{\natural}$, therefore $(1/4)$ supersymmetries are preserved even for this nontubular case.


Let us consider the radial profile of the case of $u_x\ne 0$. Inserting the above result, $B=B_0{\cal R}^2u_x$, into the Gauss law of Eq. (\ref{eom}) yields $\vert u_x\vert=E_0^{-1}$, where $E_0$ is a constant. One can always choose the solution to be $u=E_0^{-1}x$ by a suitable translation and reflection of the configuration along the axial direction. The remaining equations of motion in Eq. (\ref{eom}) require that $\mbox{sgn}(u_x)$ be independent of the coordinates $x$ and $v$. These are already satisfied by the above solution. Therefore, the supersymmetric configuration is summarized as
\begin{eqnarray}\label{sol}
x=E_0u,\quad E=1,\quad B=B_0{\cal R}^2u_x.
\end{eqnarray} 

We note here that the solution $u=E_0^{-1}x$ becomes negative for the negative value of $x$ although the elliptic coordinate system is conventionally defined for the positive value of $u$.
Since the coordinate $x$ parametrizes the axial direction of the configuration, there is no reason to truncate the solution only to the positive value of $x$. 

In order to see the physical implications of the solution, we look at the asymptotic behavior of the BI field strengths. 
The above solutions (\ref{sol}) cast them in the elliptic coordinates $\{u,v\}$ into the form, $F=E_0dt\wedge du+B_0{\cal R}^2 du\wedge dv$. In the asymptotic regions ($x\rightarrow\pm\infty$), the elliptic coordinates approach the polar coordinates since the eccentricity of the ellipse, $1\mp\tanh{u}$, vanishes  there. Indeed we can consider the following circular limit. As $u\rightarrow\pm\infty \,\,(x\rightarrow\pm\infty),$ we let $a\rightarrow 0,$ with $ae^{\pm u}=2R$ kept finite:
\begin{eqnarray}\label{asymp}
y&=a\cosh{u}\cos{v}&\approx
\frac{ae^{\pm u}}{2}\cos{v}=R\cos{v},\nonumber\\
z&=a\sinh{u}\sin{v}&\approx
\pm\frac{ae^{\pm u}}{2}\sin{v}=\pm R\sin{v}.
\end{eqnarray}
In both asymptotic regions, ${\cal R}^2\approx a^2(\pm e^{\pm u}/2)^2=R^2$ and the metric becomes;
\begin{eqnarray}
ds^2&=&a^2\left(\sinh^2{u}+\sin^2{v}\right)\left(du^2+dv^2\right)\nonumber\\
&\approx&dR^2+R^2dv^2.
\end{eqnarray}
The BI field strengths asymptotically become 
\begin{eqnarray}\label{bi}
F\approx\pm\frac{E_0}{R}dt\wedge dR\pm B_0 RdR\wedge dv \quad (x\rightarrow\pm\infty),
\end{eqnarray}
which expresses the radial Coulomb-like electric fields and uniform magnetic fields. The sign flips in the charges are due to the orientation reversal between those two asymptotic regions, which is manifest in Eq. (\ref{asymp}). This implies that those two asymptotic regions can be viewed as a D$2$-brane and a $\overline{\mbox{D}2}$-brane. They are connected to each other by an elliptic throat which is flattened at $E_0u=x=0$. Fig. 1 draws the solution (\ref{sol}) for several different values of $E_0$. 

\epsfbox{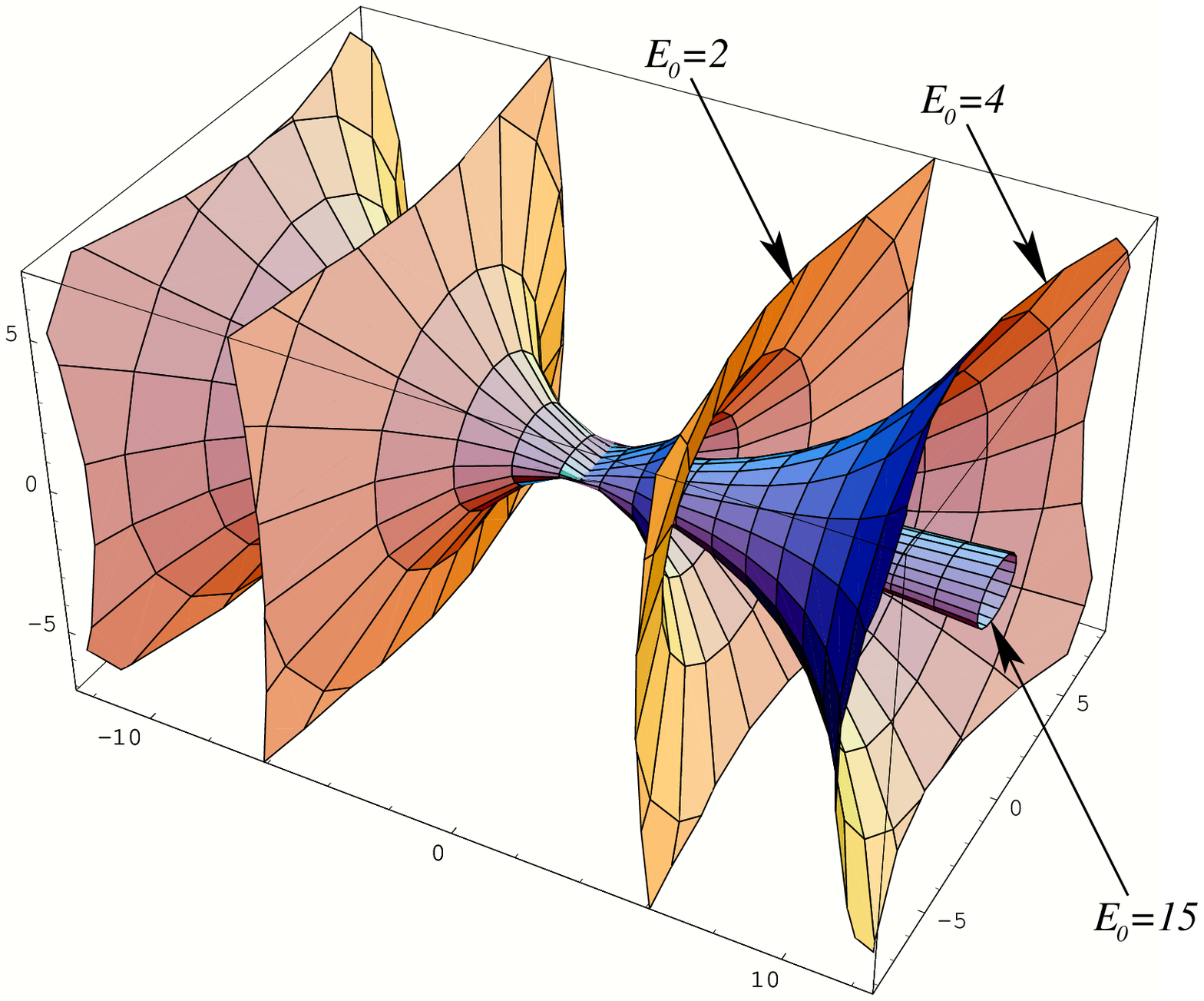}
\begin{flushleft}
{\footnotesize Fig. 1: D$2$-$\overline{\mbox{D}2}$ pairs connected by a strip of the superstrings. Three solutions are drawn for $E_0=2,4,$ and $15$ in the string unit, respectively. We set the plot range to $-12<x<12$ and the strip width, $a=1$.}
\end{flushleft}

It turns out that the elliptic throat must be a strip of type-IIA superstrings connecting a D$0$-charged D$2$-$\overline{\mbox{D}2}$ pair. Let us look at the BPS condition. Plugging the solution (\ref{sol}) into the Hamiltonian (\ref{hamiltonian}) yields the relation,
\begin{eqnarray}\label{bpsbound}
{\cal H}=\frac{\sqrt{1+B_0^2}}{B_0}B+\vert\Pi\vert.
\end{eqnarray} 
Since the solution is supersymmetric, both terms should be interpreted as the charges of some BPS objects. The second term gives the line density of the type-IIA superstrings upon the integration over the angular coordinate $v$. The same integration of the first term results in not just the line density of D$0$ particles but more than that by the factor $1/\cos{\alpha}$, which cannot be 1 for finite values of $B_0$. This implies that the BPS configuration we have found cannot be constructed via ``blowing-up" from the inhomogeneously D$0$-charged superstrings.
The correct interpretation of the first term can be obtained by its integration over the world-volume; 
\begin{eqnarray}\label{d2}
2\oint\int_0^\infty du \,\,\,{\cal R}^2\sqrt{1+B_0^2}
=2\int\int\sqrt{1+B_0^2}\,\,\,dy\wedge dz.
\end{eqnarray} 
This is just the energy sum of a D$2$- and a $\overline{\mbox{D}2}$-brane, which have the constant magnetic fields $B_0$ over their world-volumes. These BPS objects typically result from a D$2$- and a $\overline{\mbox{D}2}$-brane when D$0$ particles dissolve in them \cite{polchinski}, turning into fluxes. The net result is that the BPS configuration is rather composed of a D$0$-charged D$2$-$\overline{\mbox{D}2}$ pair and a strip of type-IIA superstrings connecting the pair.

The relation $E_0=\Pi\sqrt{1+B_0^2}$ tells us that the electric charges in Eq. (\ref{bi}) are due to the connecting superstrings and are modified by the factor $\sqrt{1+B_0^2}$ in the constant magnetic background. Geometrically this factor concerning $B_0$ can be understood as follows. 
The electric charge $E_0$ is proportional to the throat length $l$ as we see in Fig. 1. (Defining the throat region as $\{x,v\vert\,\, {\cal R}^2\lsim 1\}$, we get the relation $l=\Delta x\sim E_0$.)  The increase of the field $B_0$ confines the string end points more \cite{seiberg}, which makes the throat longer, hence the electric charge $E_0$ gets larger.


So far, we have obtained exactly a quarter-supersymmetric solution describing a D$2$-brane and a $\overline{\mbox{D}2}$-brane which are connected by an elliptically tubular throat. In the tubular case, we observed that the Poynting momentum density $\vert B\Pi\vert=C^2+a^2\sin^2{v}$ matches well with our intuition acquired from the study of the circular supertube case; high density implies large ``blowing-up.'' Near $v=0$ and $\pi$, the density is low, so the D$2$-brane curls up quite a bit there. This intuition does not fit with the nontubular case. We could not generate an arbitrarily shaped radius-varying configuration by adjusting the Poynting momentum density appropriately along the axial direction. We found that the shape is very limited to satisfy the supersymmetric condition. Conclusively, the elliptic D$2$-brane has two independent supersymmetric loci, one represented by the tubular configuration and the other represented by the radius-varying configuration. They have different BPS bound from each other. Those two loci meet on one line parametrized by the strip width, $a$. Indeed, when $B_0\rightarrow\infty$ with $E_0/B_0$ kept finite, the throat length ($\sim E_0$) becomes infinite and $u=x/E_0\rightarrow 0$ for $\vert x\vert<\infty$. This corresponds to the $C\rightarrow 0$ limit of the tubular case, hence, the flattened tube.

Different from the circular tube, the elliptic tube has a well-defined zero radius limit (with $a\ne 0$). It corresponds to the flattened D$2$-brane tube which is well described by the Lagrangian (\ref{lag}) in such a limit. This implies that given the number densities of the D$0$-branes and the superstrings, the elliptic tube stabilizes its radius to find the minimum of the Hamiltonian. This was very subtle to see in the circular tube case because the D$2$-brane description fails there.
In this regard, the shape of the tubular D$2$-brane is determined by the number densities of the D$0$-branes and the superstrings. A sufficiently large number (with $C^2\ne 0$) of the D$0$-charged superstrings forming a strip shape will be blown-up to an elliptic supertube. If the same number of the D$0$-charged superstrings is initially at one point of the $(y,\,z)$-plane, they will form a circular supertube finally. Energetically, there is no transition between those two configurations because both final states are BPS, so the net energy is the sum of those numbers of the D$0$-branes and the superstrings.   

Let us look at the M-theory origin of the nontubular configuration. The connecting superstrings come from the M2-branes winding around the compact eleventh direction. The D$2$-$\overline{\mbox{D}2}$ pair is S-dual to the M$2$-$\overline{\mbox{M}2}$ pair with their world-volume transverse to the eleventh direction. As the superstrings run from the point $x=0$, they become tubular (anti-)D2-brane acquiring another compact dimension transverse to the eleventh direction. This ``blowing-up" does not look, in the strong-coupling regime, like the ``blowing-up" of the connecting M2-branes because there is no three-dimensional BPS object in the eleven dimensions. The correct picture will be that, off the point $x=0$, the connecting M2-branes wind along the (anti) diagonal direction of the torus composed of the compact eleventh direction and the angular direction of tubular (anti-)D2-brane. An additional gravitational wave along the eleventh direction will result in a magnetic field over the (anti-)D2-brane in ten dimensions.

There is another way of understanding the nontubular configuration \cite{emparan2}. The type-IIA superstrings are under the influence of four-form RR field strength produced by a D2-brane--anti-D2-brane pair. Under this circumstance, the superstrings can tunnel into D-branes \cite{emparan}. 

Lastly, we give a brief remark on further work. Since our solution shows explicitly the relation between the geometrical deformation of the (anti-)D-branes and the potentials of the BI fields, it is tempting to study some dynamical process in this system. One could consider the scattering of a wave off the potential on either (anti-)brane. This can be done purely in the field theory setup, that is, the $(2+1)$-dimensional noncommutative $U(1)$ theory for the system at hand. The same process should have some interpretation in the D-brane language. One can check how this process can be understood geometrically. The energetic wave will propagate into the throat and transmit to other brane. Similar physics has been studied in Ref. \cite{rey} for two D$3$-branes connected by a fundamental or a D-string. 


\section*{Acknowledgments}
We appreciate Dongsu Bak, Roberto Emparan, Seungjoon Hyun, Youngjai Kiem, Yoonbai Kim, David Mateos, and Soo-Jong Rey for their helpful comments and stimulating discussions.
This work is supported in part by KOSEF through Project No. 2000-1-11200-000-3.


\begin{thebibliography}{99}
\bibitem{townsend1} The BIon spike solution was obtained in Ref. \cite{townsend} as the radius-varying circular D$2$-brane. Since it carries nonvanishing D$2$-brane charge, we rule out the model from our consideration.
\bibitem{emparan2} We owe this nice observation to Roberto Emparan.
\end{thebibliography}
\end{document}